\documentclass{aa}

\usepackage{graphicx}
\usepackage[varg]{txfonts}

\usepackage{adjustbox}
\usepackage{arydshln}

\newcommand{\GG}[1]{}
\defcitealias{Bh+19}{Paper I}
\defcitealias{Bh+19b}{Paper II}

\begin{document}

\title{The survey of planetary nebulae in Andromeda (M~31)} \subtitle{III. Constraints from deep planetary nebula luminosity functions on the origin of the inner halo substructures in M~31} 
    \titlerunning{The survey of PNe in M 31 III - Substructure origin from deep PNLF}
    \author{Souradeep Bhattacharya\inst{1}\and
          Magda Arnaboldi\inst{1} \and 
          Ortwin Gerhard\inst{2} \and
          Alan McConnachie\inst{3} \and
          Nelson Caldwell\inst{4} \and
          Johanna Hartke\inst{5} \and
          Kenneth C. Freeman\inst{6}
          }
   \institute{European Southern Observatory, Karl-Schwarzschild-Str. 2, 85748 Garching, Germany \\ 
   \email{sbhattac@eso.org} \and
   Max-Planck-Institut für extraterrestrische Physik, Giessenbachstraße, 85748 Garching, Germany \and
   NRC Herzberg Institute of Astrophysics, 5071 West Saanich Road, Victoria, BC V9E 2E7, Canada \and
   Harvard-Smithsonian Center for Astrophysics, 60 Garden Street, Cambridge, MA 02138, USA \and
   European Southern Observatory, Alonso de C\'ordova 3107, Santiago de Chile, Chile \and   
   Research School of Astronomy and Astrophysics, Mount Stromlo Observatory, Cotter Road, ACT 2611 Weston Creek, Australia 
             }

   \date{Submitted: May, 2020; Accepted: January, 2021}

 
  \abstract
   {The Andromeda (M~31) galaxy displays several substructures in its inner halo. Different simulations associate their origin with either a single relatively massive merger, or with a larger number of distinct, less massive accretions.}
   { The origin of these substructures as remnants of accreted satellites or perturbations of the pre-existing disc would be encoded in the properties of their stellar populations (SPs). The metallicity and star formation history of these distinct populations leave traces on their deep [\ion{O}{iii}] 5007$\AA$ planetary nebulae luminosity function (PNLF). By characterizing the morphology of the PNLFs, we constrain their origin.}
   {From our 54 sq. deg. deep narrow band [\ion{O}{iii}] survey of M~31, we identify planetary nebulae (PNe) in six major inner-halo substructures -- the Giant Stream, North East Shelf, G1-Clump, Northern Clump, Western Shelf and Stream-D. We obtain their PNLFs and those in two disc annuli, with galactocentric radii R$\rm_{GC}$=10--20 kpc and R$\rm_{GC}$=20--30 kpc. {We measure PNLF parameters from cumulative fits and statistically compare the PNLFs in each substructure and disc annulus.} We link these deep PNLF parameters and those for the Large Magellanic Cloud to published metallicities and { resolved stellar population-age measurements} for their parent SPs.}
   {{The absolute magnitudes ($M^{*}$) of the PNLF bright cut-off for these sub-populations span a significant magnitude range, despite being located at the same distance and having a similar line-of-sight extinction. The $M^{*}$ values of the Giant Stream, W-shelf and Stream-D PNLFs are fainter than those predicted by PN evolution models by 0.6, 0.8, and 1.5 mag, respectively, assuming the measured metallicity of the parent stellar populations. The faint-end slope of the PNLF increases linearly with decreasing fraction of stellar mass younger than 5 Gyr across the M~31 regions and the LMC. From their PNLFs, the Giant Stream and NE-shelf are consistent with being stellar debris from an infalling satellite, while the G1 Clump appears to be linked with the pre-merger disc with an additional contribution by younger stars. }
   }
   {The SPs of the substructures are consistent with those predicted by simulations of a single fairly massive merger event that took place 2--3 Gyr ago in M31. Stream-D has an unrelated, distinct, origin. Furthermore, this study provides independent evidence that the faint-end of the PNLF is preferentially populated by planetary nebulae evolved from older stars.}


\maketitle

%

\begin{figure*}[t]
	\centering
	\includegraphics[width=\textwidth,angle=0]{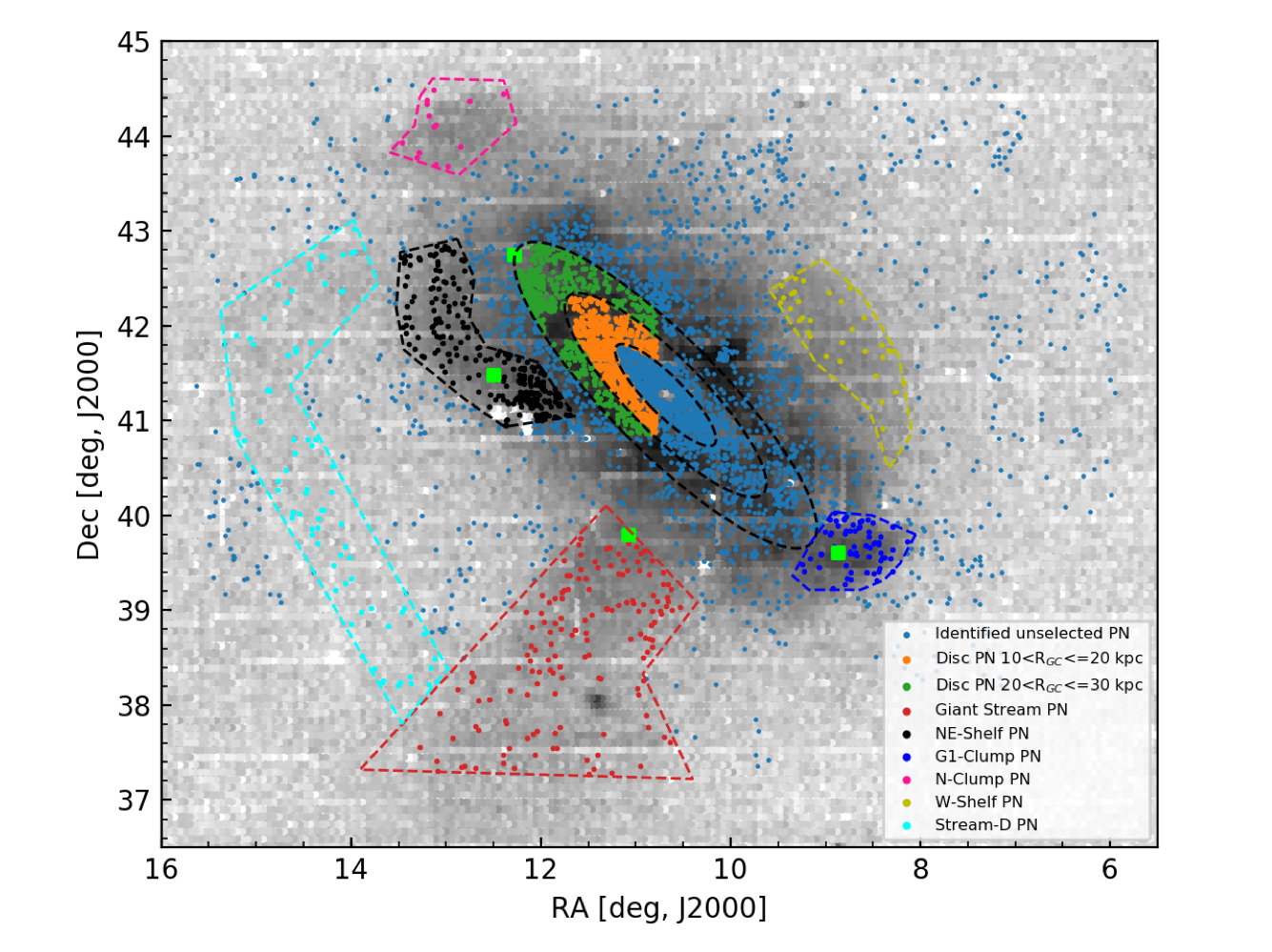}
	\caption{Position on sky of PNe (all marked points) identified by \citetalias{Bh+19} and Bhattacharya et al. (2021, in preparation). They are overlaid on the number density map of RGB stars from PAndAS \citep{mcc18}, binned for visual clarity and shown in grey. The PNe selected for the PNLF analysis in the two disc regions and six substructure regions are marked with different colours. North is up, {east is left.} The HST fields observed by \citet{bernard15} to obtain the SFH of four of the M~31 regions have been marked with green squares.}
	\label{fig:spat}
\end{figure*}

\section{Introduction}
Within the Lambda cold dark matter ($\Lambda$CDM) cosmological model, galaxies evolve by hierarchical mass accretion \citep{White78,Bullock05}. Fossil records of {these processes} are observed in the outskirts of massive galaxies in the form of shells, streams, tidal tails and other substructures \citep[e.g.][]{Mihos05,MartinezDelgado10,Crnojevic16}. The masses and chemical properties of the accreted satellites may be uncovered from the kinematics and chemical abundances of surviving substructures \citep{Johnston08}.  Andromeda (M~31) is the nearest giant spiral galaxy to the Milky Way (MW) at a distance of $\sim773$ kpc \citep{Conn16}. The Pan-Andromeda Archaeological Survey \citep[PAndAS;][]{mcc09,mcc18} produced a map of resolved red giant branch (RGB) star counts that revealed {multiple} substructures (Figure~\ref{fig:spat}) out to a projected radii of $\sim150$ kpc around M~31. These features unveiled the tumultuous merger history of M~31. With faint structures detected down to unprecedented surface brightness levels ($\mu_{V}\sim33$mag arcsec$^{-2}$), M~31 provides the best current laboratory for studying the signatures of extended galaxy assembly.

The Giant Stream is the most prominent substructure in the M~31 halo. It was first discovered by \citet{Ibata01} who linked its origin to that of an infalling satellite. The subsequent discoveries of other substructures, particularly those near the disc-halo interface, namely -- North East Shelf (NE-Shelf), G1-Clump, Northern Clump (N-Clump), Western Shelf (W-Shelf) and Stream-D, led to intense speculations on their origin. \citet{Fardal13} showed that N-body simulations of a satellite of $\rm \sim 3.2 \times 10^{9} M_{\odot}$ infalling along the Giant Stream $\sim1$ Gyr ago can reproduce the structure and kinematics of the Giant Stream and also of the NE and W shelves.

The N and G1-Clump, though, do not arise naturally in models of the dissolution of the Giant Stream progenitors from minor mergers, and they were linked to distinct accretion events of dwarf galaxies  \citep{fm16}. 
Most recently, using hydrodynamical simulations, \citet{ham18} argued that a single merger of a more massive satellite  with total mass $1.4\times10^{10}M_\odot$, occurring 2 -- 3 Gyr ago, is able to perturb the pre-existing M31 disc leading to the origin of the substructures along its major axis, as well as that of the Giant stream and the associated NE and W shelves.  

The morphology of the low surface brightness substructures, the $2$ Gyr old burst of star formation, observed in the M~31 disc and in some inner halo \textit{Hubble} Space Telescope (HST) pointings \citep{bernard15,wil17}, plus the dynamical heating of the disc  \citep[][hereafter Paper II]{dorman15,Bh+19b} are in line with the predictions from a $1:4\, -\, 1:5$ mass merger in M~31 \citep{ham18}. However, further investigation is required to link the origin of any of the observed substructures with either the pre-existing disc or the satellite remnant. 

The HST observations by \citet{Richardson08} and subsequent star formation histories (SFHs) computed by \citet{bernard15} for 14 pencil beam pointings, some of which are aligned with the M~31 overdensities having stellar populations similar to that of the Giant Stream while others are more similar to the M~31 disc. An intrinsic limit of this approach though is the very small area coverage of single HST pointings (see Figure~\ref{fig:spat}) within a significantly more extended sub-region in the M~31 inner halo \citep{wil17}. Therefore, the SFH measured in such a small area, ``localized-SFH''' hereafter, may not provide a robust representation of the SFH over the entire extended sub-region if there is a significant variance/gradient of the stellar population properties; see for example, the metallicity gradient observed within the Giant Stream \citep{Conn16}. 
In this study we then turn to the deep luminosity function of the associated Planetary Nebulae (PNe) population, which provides few constraints on the SFH, but in return these are obtained for an entire subregion.

PNe are identified from their relatively strong [\ion{O}{iii}] 5007$\AA$ emission and negligible continuum. The characteristic [\ion{O}{iii}] 5007$\AA$ (M$\rm_{5007}$) PN luminosity function (PNLF) was first described by \citet{ciardullo89}. The analytical formula
\begin{equation}
    \rm N(M) \propto e^{0.307M}(1-e^{3(M^{*}-M)})
\end{equation}
was fitted to the bright PNe detected in the centre of M 31 
\citep{ciardullo89}. 
The PNLF is a reliable secondary distance indicator for determining galactic distances out to $\sim$20 Mpc by virtue of its invariant absolute bright cut-off, M$^{*}$, currently measured at $\rm M^{*}_{ref}= -4.54 \pm 0.05$ at near-solar metallicities \citep{ciardullo13}. While M$^{*}$ gets fainter in low metallicity populations \citep[e.g.][]{ciardullo92,ciardullo02,hmp09}, according to theoretical predictions by \citet{Dopita92}, it seems to be invariant {to} either the age of the population or galaxy morphological type. The faint end of the PNLF is theoretically expected to follow an exponential function \citep{jacoby80} describing the slow evolution of the central star (CSPN) embedded in a rapidly expanding, optically thin nebula \citep{hw63}. 

The slope of the PNLF measured at a few magnitudes fainter than its bright cut-off is observed to vary depending on the SFH of the average parent stellar population. In galaxies where the PNLF has been observed to $\sim$3 mag below M*, steeper PNLF slopes were observed for elliptical galaxies with older stellar populations \citep[e.g.][]{longobardi13, hartke17} and conversely flatter slopes for younger populations \citep[e.g. M33;][]{ciardullo04}. 
In surveys deeper than $\sim$3 mag below M*, the shape of the PNLF may exhibit further changes in slope, as seen in the Large Magellanic Cloud \citep[LMC;][]{rp10} or dips, e.g. that in the Small Magellanic Cloud \citep[SMC;][]{jd02} and NGC~6822 \citep[a Local Group dwarf irregular galaxy;][]{hmp09}. \citet{rod15} describe a two-mode PNLF for NGC~6822 and show that these modes of the PNLF may be associated with the two episodes of star formation, with the younger and older parent stellar population dominating the brighter and fainter PNe respectively. Indeed \citet{Valenzuela19} showed theoretically that the shape of the PNLF at the faint-end is quite sensitive to the distribution of masses of the PN central stars, which is a product of the SFH in any galaxy. In \citet[][hereafter Paper I]{Bh+19}, we observed a rise in the faint-end of the PNLF in the central 16 sq. deg. of M~31 conceivably linked to the SFH of M~31.

Owing to the homogeneous spatial coverage and  photometry of the current imaging survey, we detect PNe associated with all the stellar populations in a 54 sq. deg area centred on M31. Thanks to the sensitivity and accuracy, we can quantify the imprints of the variations of these stellar populations on the PNLFs, in distinct regions. In Section~\ref{sect:data}, we describe our PN survey of M~31 and set the geometrical boundaries around any regions with RGB overdensities where the PNLFs are measured. In Section~\ref{sect:pnlf}, we obtain the PNLFs, compute the parameters that best describe them and carry out the quantitative comparisons via  statistical methods.  In Section~\ref{sect:pnlf_reg}, we correlate the stellar population parameters with the properties of the PNLFs at both their bright and faint ends in the different subregions. In Section~\ref{sect:disc}, we discuss the implications of our results on the origin of the M31 inner halo substructures. We draw our conclusions in Section~\ref{sect:future}.

\begin{table}[t]
\caption{No. of PNe in each studied region in our M 31 survey.}
\centering
\adjustbox{max width=\columnwidth}{
\begin{tabular}{c|c|c}
\hline
Region & N$\rm_{PN}$ & N'$\rm_{PN}$\\
\hline
Disc (10<R$\rm_{GC} \leq $20 kpc) & 705 &  624\\
Disc (20<R$\rm_{GC} \leq $30 kpc) & 507 &  321\\
Giant Stream & 151 &  113\\
NE-Shelf & 250 &  106\\
G1-Clump & 65 &  48\\
N-Clump & 14 &  8\\
W-Shelf & 32 &  31\\
Stream-D & 78 &  36\\
\hline
\end{tabular}
\label{table : obs}
}
\tablefoot{ {N$\rm_{PN}$-- Total number of PNe identified in each M~31 region; N'$\rm_{PN}$-- Number of PNe identified within the magnitude range m$\rm_{5007}\leq $26.16 mag in each M~31 region.} }
\end{table}


\section{Survey of Planetary Nebulae in M31}
\label{sect:data}

\subsection{Survey characteristics}
\label{sect:surv_car}
In \citetalias{Bh+19}, we identified PNe candidates in a 16 sq. deg. uniform imaging survey of the disc and disc-halo interface of M~31 using the MegaCam wide-field imager \citep{boulade03} mounted on the 3.6-meter Canada-France-Hawaii Telescope (CFHT). M~31 was observed through a narrow-band [\ion{O}{iii}] filter ($\lambda\rm_c = 5007~\AA, ~\Delta\lambda = 102~\AA$, on-band) and a broad-band \textit{g}-filter ($\lambda\rm_c = 4750~\AA, ~\Delta\lambda = 1540~\AA$, off-band). Additional 38 sq. deg. were obtained in late-2019, providing uniform coverage of the inner halo of M~31 and its substructures reaching a uniform sensitivity down to fluxes equivalent to m$\rm_{5007} = 26.2$ mag over the entire area. Details of these observations are presented in \citet{mythesis} and Bhattacharya et al. (2021, in preparation). 

The zero points for each of the [\ion{O}{iii}] and \textit{g}-band frames in \textit{AB} magnitudes, normalised to a 1 s exposure, are Z$_{\rm[\ion{O}{iii}]}$ = 23.4 and Z$_g$ = 26.5. The photometry was calibrated using  observations of spectrophotometric standard stars. The photometric uncertainty (err$\rm_{m,5007}$) is well below 0.01 mag at m$\rm_{5007} = 22.5$ mag; it increases to err$\rm_{m,5007}\sim$0.15--0.2 mag at m$\rm_{5007} = 26.2$ mag.

The PN identification in each of these pointings was also carried out as described in \citetalias{Bh+19} using the on-off band technique developed and validated in \citet{arnaboldi02,arnaboldi03}, later optimised for large imaging surveys by \citet{longobardi13}. The PNe are identified as point-like sources having an excess [\ion{O}{iii}] - \textit{g} colour. The complete imaging survey led to the identification of 5265 PN {candidates} in M~31, the largest PN sample in any galaxy ({\citetalias{Bh+19}}; Bhattacharya et al. 2021, in preparation). Of these, 4085 are newly discovered, down to m$\rm_{5007} \leq 26.7$ mag in the deepest pointing. In \citetalias{Bh+19}, we verified that the identified PNe have counterparts in the HST data from the Panchromatic \textit{Hubble} Andromeda Treasury \citep[PHAT;][]{dal12} in the regions of overlap, with at most $\sim3\%$ possible contamination.

\subsection{Substructures in M~31 and their PN sub-samples}
In Figure~\ref{fig:spat} we show all the PNe (all coloured points) identified by the narrow band imaging survey. They are overlaid on the map of RGB stars from PAndAS \citep{mcc18}. The prominent substructures appear as overdensities in this map of RGB stars. The distances to any of the M~31 substructures are consistent with that of the disc \citep[773 kpc;][]{Conn16}. Akin to \citet{mcc18}, we define the spatial boundaries of these substructures with polygons around the regions of higher spatial density. The identified overdensities, as marked with dashed lines in Figure~\ref{fig:spat}, are the Giant Stream (red), North-East Shelf (NE-Shelf; black), G1 Clump (dark blue), Northern Clump (N-Clump; pink), Western Shelf (W-Shelf; yellow) and Stream-D (cyan). The PNe included within a given region are considered to belong to that substructure and used for subsequent analysis. Figure~\ref{fig:spat} also shows elliptical projections of circular annuli in the disc (black dashed) at galactocentric distances of R$\rm_{GC}$= 10, 20 and 30 kpc, covering the disc of M~31. As found in \citetalias{Bh+19b}, the PNe in the 10<R$\rm_{GC} \leq $20 kpc disc region of M~31 belong both to the (heated) thin and thicker disc, while the PNe in the 20<R$\rm_{GC} \leq $30 kpc disc region are dominated by PNe belonging to the thicker disc of M~31. 

In subsequent sections, we only consider PNe in the disc regions which were selected from the fields with limiting magnitude of m$\rm_{5007}\geq $26.16 mag\footnote{The fields in the {south-west} of the M~31 disc have a brighter limiting magnitude (50\% detection completeness limit computed from injection of artificial sources) of m$\rm_{5007}\sim$25.89 mag. We thus confine the analysis of the PNLFs in the disc to the fields for the {north-east} corner of the M~31 disc with sensitivity as deep as that reached in the substructures' fields. See \citetalias{Bh+19} for details of limiting magnitudes of different fields.}. Figure~\ref{fig:spat} shows the PNe in the defined 10<R$\rm_{GC} \leq $20 and 20<R$\rm_{GC} \leq $30 kpc disc regions in orange and green respectively. Table~\ref{table : obs} lists the different regions studied in this work along with the number of PNe identified in those regions, N$\rm_{PN}$, and those within the magnitude range m$\rm_{5007}\leq $26.16 mag utilized in the subsequent analysis, N'$\rm_{PN}$. 

\subsection{Extinction in the M31 survey region}
\label{sect:ext_sur}

The line-of-sight (LOS) extinction to any subregion in M31 depends on the extinction from the MW halo \citep{Schlegel98} and on the internal extinction in M~31 and its substructures. As shown by \citet{ibata14} for the PAndAS survey region (see their Fig.~2) which overlaps the area in M~31 covered by this work, the MW halo LOS extinction is largely homogeneous having almost the same value (A$\rm_{V}\sim0.19$ mag) for the entire M31 disc and its substructures. {The same can also be observed in the latest foreground reddening map obtained from LAMOST spectra of the MW halo stars to the M31 LOS \citep[][see their Figure 2]{Zhang20}.}
 
Regarding the extinction by dust internal to M31, the dust surface density map of \citet{Draine14} shows a near-exponential radial decrease in the M31 disc, with a scale-length of a few kpc outside an annular radius of $R_{gc}=15$ kpc. Combined with the dust model of \cite{Draine07}, this corresponds to an average A$\rm_{V}\sim0.3$ mag along the line-of-sight at $R_{gc}=20$ kpc and suggests negligible values beyond $R_{gc}=30$ kpc where the inner halo substructures are located (see Fig.~\ref{fig:spat}). 

Therefore dust extinction is present mostly within the $R_{gc}=20$ kpc elliptical isodensity contour of the M31 disc, thereby mainly affecting the PN population in the 10-20 kpc disc region {considered} in this work. If most of the dust is located together with the gas at very low scale heights, the PNe identified in the 10-20 kpc disc region might miss a fraction of PNe behind the dust layer along the LOS, but the bright cut off would still reflect the cut off of the bright PNe in front of the dust.

Outside this annulus, the effect of dust internal to M31 is small in the 20--30 kpc disc region and negligible in the substructures further out. From the PAndAS CMDs \citep{ibata14,mcc18} and the CMDs of the small HST fields by \citet{bernard15} at $R_{gc} \geq 30$ kpc, there is no evidence for differential reddening. From the recent LAMOST extinction map, the reddening caused by dust in the M31 halo decreases from about 0.01 mag in the central regions to
0.001 mag at a projected distance of 5 R$_{25}$ ($\sim 100$ kpc; \cite{Zhang20}). Hence, no further correction is required for internal extinction in any regions other than the 10--20 kpc {annulus}.

In the subsequent sections, we consider the LOS extinction of A$\rm_{5007}\sim0.21$ mag {(corresponding to A$\rm_{V}\sim0.19$; \citealt{Cardelli89})} for the M~31 disc and each of the substructures. We note that circumstellar extinction of individual PN is their intrinsic property, linked to their progenitor (see Appendix~\ref{sect:ext} for details). Hence, circumstellar extinction is not to be accounted for whenever deriving the absolute magnitude of the PNLF bright cut off.


\begin{figure}[t]
	\centering
	\includegraphics[width=\columnwidth,angle=0]{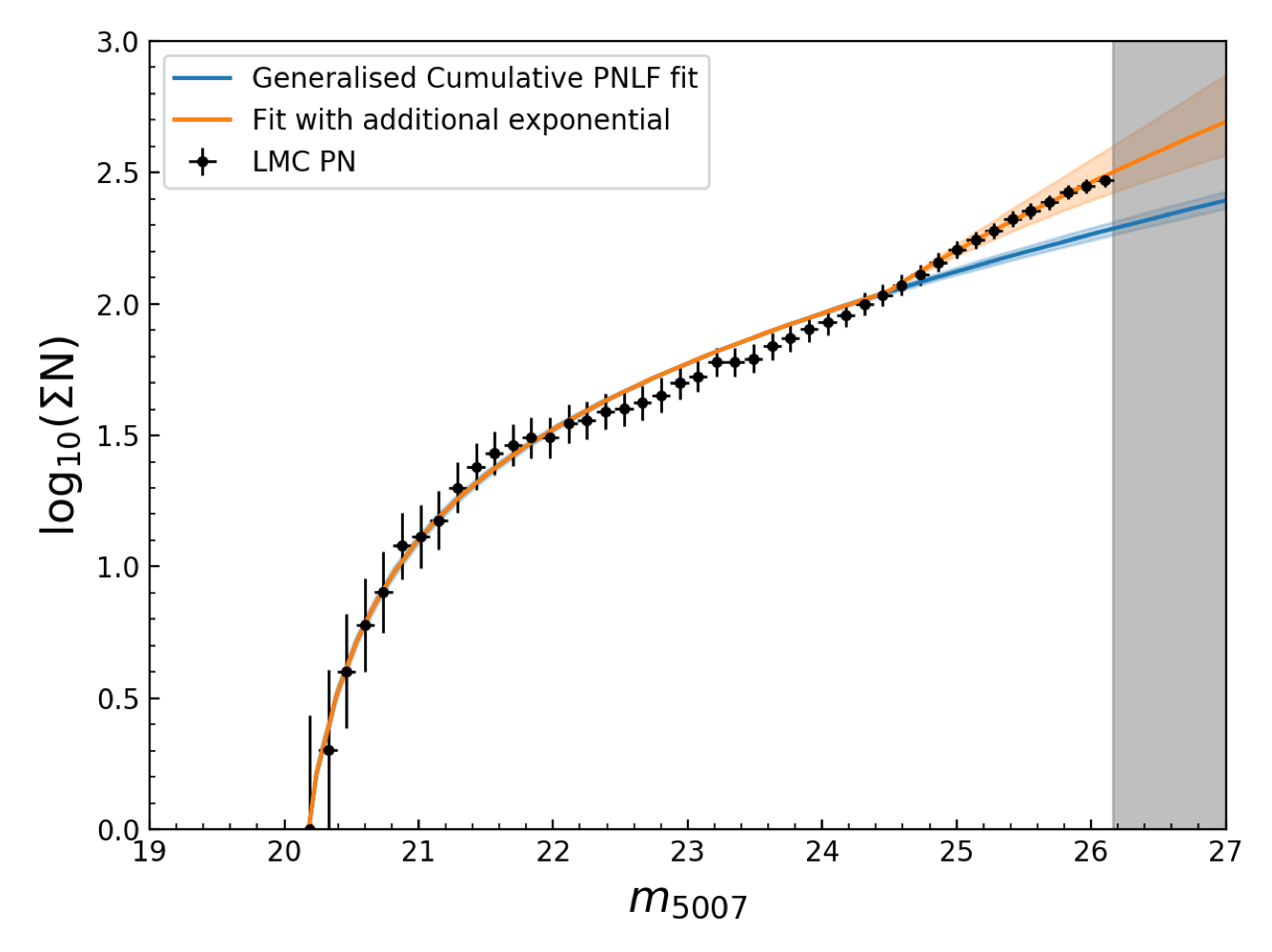}
	\caption{Cumulative PNLF for the LMC PN sample from \citet{rp10} shifted to the apparent bright cut-off of the M 31 disc region at 10<R$\rm_{GC} \leq $20 kpc. The cumulative PNLF are fitted by the generalised cumulative function (in blue) for m$\rm_{5007}\leq $24.5 mag and a function which additionally includes an exponential at the faint-end (in orange).  The uncertainty in the fits are shaded. The region fainter than the limiting magnitude of the shallowest field (m$\rm_{5007}$>26.16 mag) in our M~31 survey is shown in grey.}
	\label{fig:pnlf_cum_lmc}
\end{figure}

\begin{table*}[htb]
\centering
\caption{{Cumulative PNLF fit parameters and corresponding region properties.}}
\adjustbox{max width=\textwidth}{
\begin{tabular}{c|c|c|c|c|c|c}
\hline
Region & $M^{*}$ & c$_1$ & c$_2$ & c$_{f1}$ & c$_{f2}$ & [M/H]\\
\hline
LMC & -4.44 $\pm$ 0.00 &  40.49 $\pm$ 3.10 & 0.22 $\pm$ 0.03 & 43.95 $\pm$ 9.38 & 0.56 $\pm$ 0.31 & -0.61 $\pm$ 0.05 \\
Disc (10-20 kpc) & -4.55 $\pm$ 0.03 & 78.73 $\pm$ 5.49 & 0.27 $\pm$ 0.03 & 0.31 $\pm$ 0.38 & 5.88 $\pm$ 0.94  & -0.2 $\pm$ 0.07\\
Disc (20-30 kpc) & -4.25 $\pm$ 0.05 & 29.96 $\pm$ 2.05 &  0.23 $\pm$ 0.03 & 0.23 $\pm$ 0.03 & 8.00 $\pm$ 1.68  & -0.45 $\pm$ 0.18 \\
Giant Stream & -3.60 $\pm$ 0.17 & 8.33 $\pm$ 1.26 & 0.32 $\pm$ 0.10 & 0.03 $\pm$ 0.06 & 6.67 $\pm$ 1.40  & -0.52 $\pm$ 0.17 \\
NE-Shelf & -4.59 $\pm$ 0.11 & 3.38 $\pm$ 0.29 & 0.00 $\pm$ 0.04 & 0.01 $\pm$ 0.01 & 7.56 $\pm$ 0.40  & -0.4 $\pm$ 0.1\\
G1-Clump & -4.83 $\pm$ 0.25 & 2.80 $\pm$ 0.57 & 0.04 $\pm$ 0.10 & 0.49 $\pm$ 0.23 & 4.01 $\pm$ 0.39  & -0.37 $\pm$ 0.06\\
N-Clump & -5.92 $\pm$ 0.39 & 0.83 $\pm$ 0.08 & 0.00 $\pm$ 0.00 & N.A. & N.A. & N.A.  \\
W-Shelf & -3.40 $\pm$ 0.12 & 3.89 $\pm$ 0.79 & 0.01 $\pm$ 0.13 & 0.01 $\pm$ 0.01 & 9.96 $\pm$ 7.59 & -0.52  $\pm$ 0.09\\
Stream-D & -2.69 $\pm$ 0.18 & 1.83 $\pm$ 0.21 & 0.00 $\pm$ 0.05 & 0.03 $\pm$ 0.11 & 4.82 $\pm$ 1.58 & -1.10 $\pm$ 0.25 \\
\hline
\end{tabular}
\label{table : pnlf}
}
\tablefoot{{The cumulative PNLF fit parameters in the different regions are shown in Columns 2--6. The faint-end exponential could not be constrained for the N-Clump. The last column lists the [M/H] measurement corresponding to each region which, along with the relevant literature references, are discussed in Appendix~\ref{app:met}}.}
\end{table*}

\section{The Planetary Nebula Luminosity Function}
\label{sect:pnlf}

\subsection{Fitting the Cumulative PNLF}
\label{sect:pnlf_cum}

A quantitative analysis of the luminosity function of the PN subsamples selected in any M~31 region can be performed on the cumulative luminosity function to avoid potential histogram binning issues such as the bin size or position of the first magnitude bin \citep[e.g.][for NGC 3109]{pena07}. For an observed PN population (as in Sections \ref{sect:pnlf_lmc} \& \ref{sect:pnlf_m31}), the cumulative PNLF is constructed by taking into account the detection and selection completeness. The detection completeness correction accounts for the non-detection of PNe due to noise.  Colour or point-like selection criteria which would exclude those PNe affected by photometric errors is accounted for with the selection completeness correction. See \citetalias{Bh+19} for further detailed information. 

The PNLFs for PN populations in different galaxies are described by the generalised analytical formula introduced by \citet{longobardi13}, which is given hereafter:
\begin{equation}
$$\rm N(M)=c_1e^{c_2M}(1-e^{3(M^{*}-M)})$$
\end{equation}
where $c_1$ is a normalisation constant, $c_2$ is the slope { in the intermediate magnitude range}. The \citet{ciardullo89} LF is then a specific case of the generalised formula with $c_2$ = 0.307. The cumulative PNLF corresponding to the generalised analytical PNLF \citepalias{Bh+19} is:
\begin{equation}
$$\rm I(M)=c_1e^{c_2M}[\frac{1}{c_2}e^{c_2M}+\frac{1}{3-c_2}e^{3(M^{*}+\mu)-(3-c_2)M}-(\frac{1}{c_2}+\frac{1}{3-c_2})e^{c_2(M^{*}+\mu)}]$$
\label{cum_brt}
\end{equation}
Its free parameters are $c_1$, $c_2$ and $M^{*}$. $M^{*}$ is theoretically expected to become fainter with decreasing metallicity \citep{Dopita92}. 

In \citetalias{Bh+19}, we found an additional ubiquitous rise in the faint-end of the M~31 PNLF, with  at apparent m$\rm_{5007}$ magnitudes fainter than m$\rm_{5007}= $25 mag. We describe such rise with an additional exponential function in the cumulative PNLF which has the following form: 
\begin{equation}
$$\rm I_{f}(M) =\frac{c_{f1}}{c_{f2}}(e^{c_{f2}M}-1)$$
\end{equation}
where $c_{f1}$ is a normalization constant and $c_{f2}$ is the slope of this exponential function. Thus the cumulative PNLF can be described over the entire magnitude range by:
\begin{equation}
$$\rm I_{\rm tot}(M) = I(M)+I_{f}(M)$$
\label{cumtot}
\end{equation}
For a given PN population, the  free parameters in Equation~\ref{cum_brt} are fitted to the PN brighter than m$\rm_{5007}=25$ mag\footnote{The free parameters in Equation~\ref{cum_brt} are fitted to the PN brighter than { m$\rm_{5007}=25$ mag for the M~31 subregions, and  m$\rm_{5007}=24.5$ mag for the LMC (see  description in Section~\ref{sect:pnlf_lmc}), beyond which their respective PNLFs show the rise at the faint-end.}}. Once these are determined, they are kept fixed, and the additional parameters, $c_{f1}$ and $c_{f2}$, for $I_{f}(M)$ in Equation~\ref{cumtot} are determined from the PNe with m$\rm_{5007}\leq $26.04 mag. {At the distance and LOS extinction to the main disc of M31, the apparent magnitude of the PNLF bright cut-off is m$\rm_{5007}= $20.16 mag.}

\subsection{Independent calibration to a large PN sample- Cumulative PNLF of the LMC}
\label{sect:pnlf_lmc}

\begin{figure*}[t]
	\centering
	\includegraphics[width=\columnwidth,angle=0]{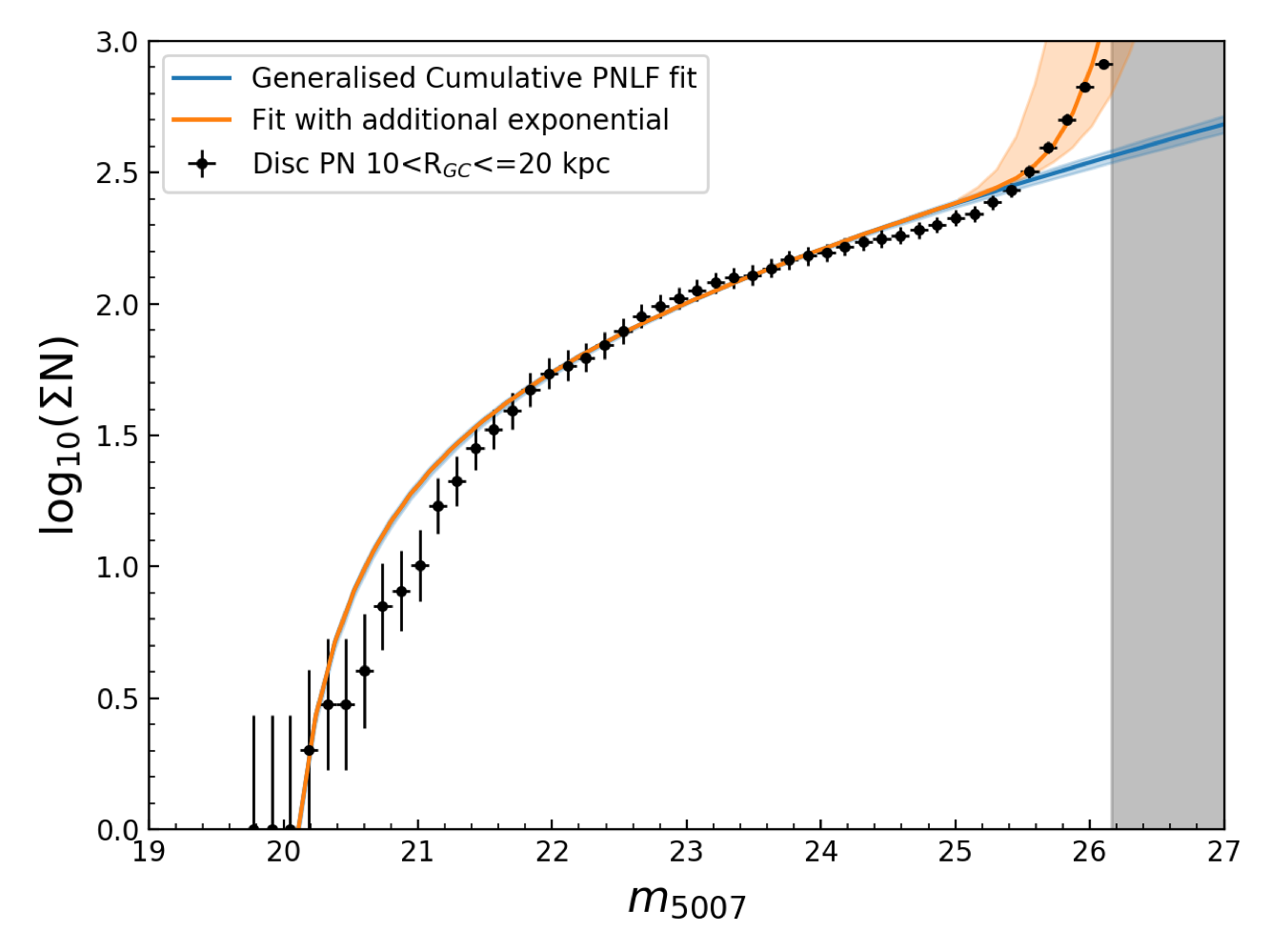}
	\includegraphics[width=\columnwidth,angle=0]{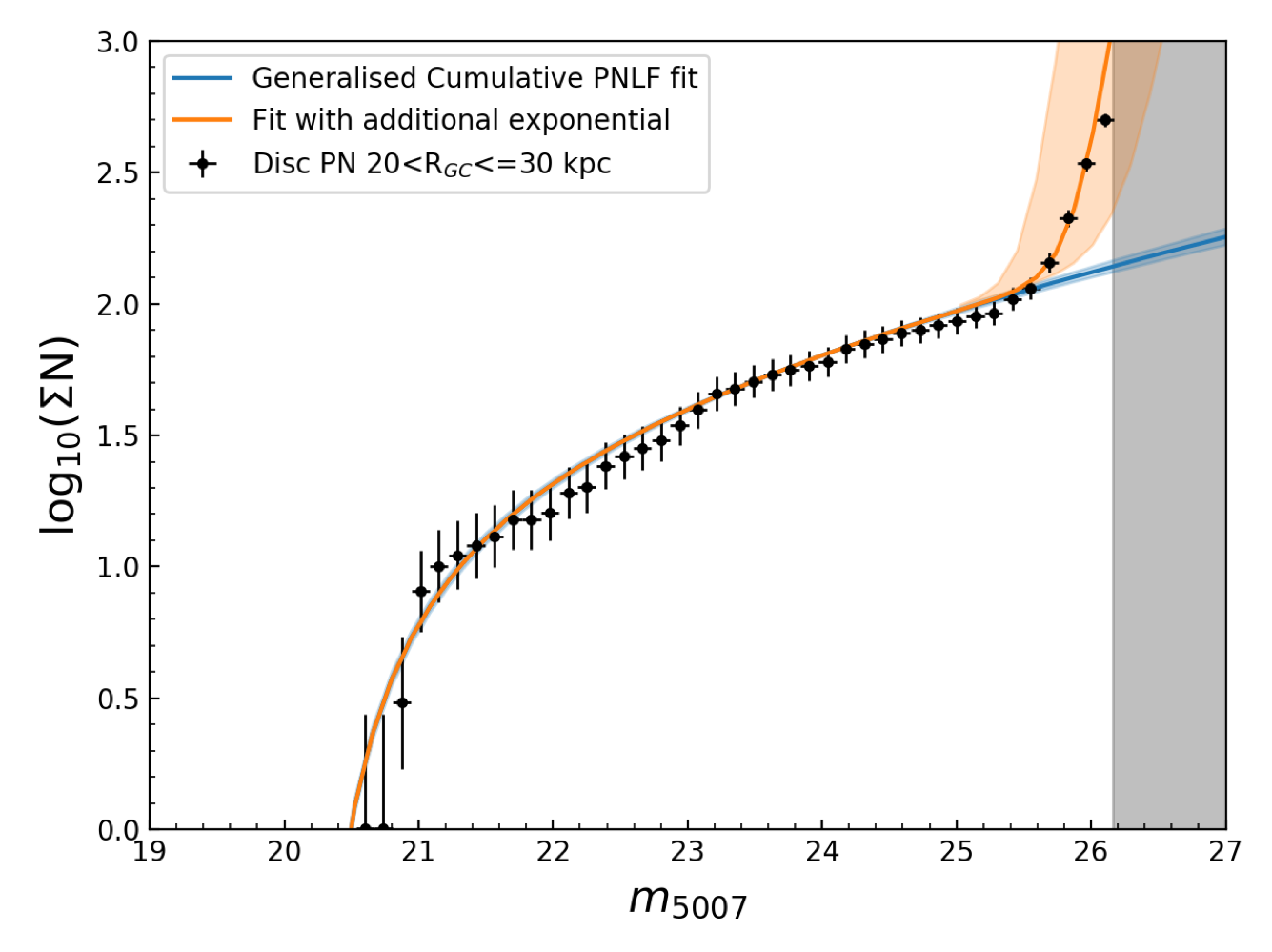}
	\caption{Completeness-corrected cumulative PNLF for the disc regions within 10<R$\rm_{GC} \leq $20 kpc (left) and 20<R$\rm_{GC} \leq $30 kpc (right). The cumulative PNLF are fitted by the generalised cumulative function (in blue) for m$\rm_{5007}\leq $25 mag and a function which additionally includes an exponential at the faint-end (in orange). The uncertainty in the fits are shaded. The region fainter than the limiting magnitude of the shallowest field (m$\rm_{5007}$>26.16 mag) is shown in grey.}
	\label{fig:pnlf_cum_disc}
\end{figure*}

\begin{figure*}[t]
	\centering
	\includegraphics[width=\textwidth,angle=0]{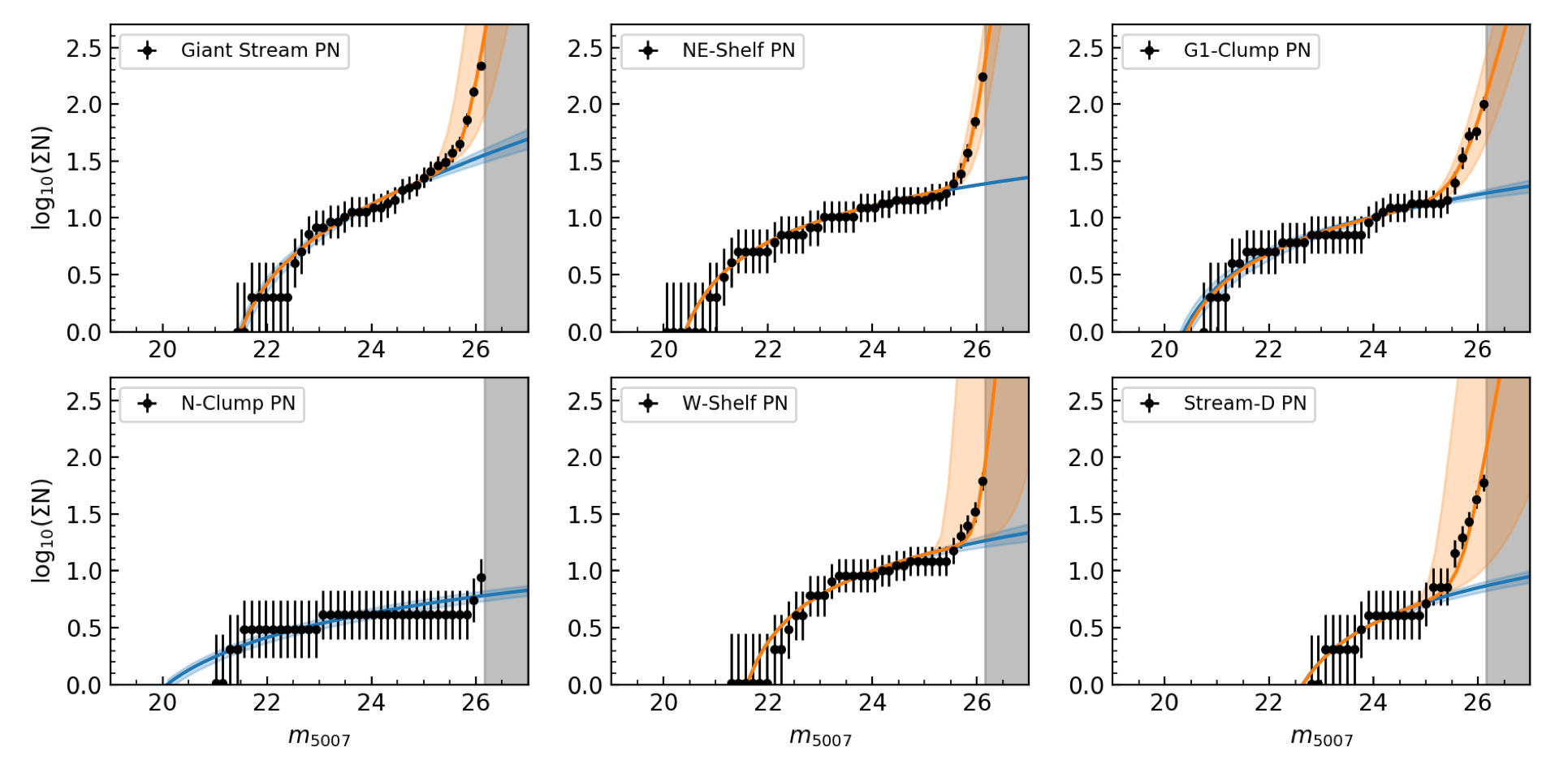}
	\caption{Completeness-corrected cumulative PNLF for the six different substructures are shown. The cumulative PNLF are fitted by the generalised cumulative function (in blue) for m$\rm_{5007}\leq $25 mag and a function which additionally includes an exponential at the faint-end (in orange; except for N-Clump). The uncertainty in the fits are shaded. The region fainter than the limiting magnitude of the shallowest field (m$\rm_{5007}$>26.16 mag) is shown in grey.}
	\label{fig:pnlf_cum_subs}
\end{figure*}

\citet{rp10} measured the PNLF in the LMC reaching $\sim8$ mag fainter than the bright cut-off. This is one of the deepest survey of PNe in any galaxy, making it a suitable independent data-set to apply our fitting procedure. By keeping the absolute bright cut-off of the more metal-poor LMC PNLF constant, $M^{*}=-4.44$ \citep{rp10}, we obtain the cumulative PNLF of the LMC (Figure~\ref{fig:pnlf_cum_lmc}) shifted to the distance and average foreground extinction  of M~31. It is fitted as previously described to derive the parameter values noted in Table~\ref{table : pnlf}. The LMC PNLF is also well described by the $I_{\rm tot}(M)$ function; the faint-end of the LMC cumulative PNLF indeed requires the additional exponential component to reproduce the rise at the faint magnitudes.

\subsection{Cumulative PNLFs of the M~31 regions}
\label{sect:pnlf_m31}

We obtain the cumulative PNLF in each identified region of M~31, having corrected for the average foreground extinction. Each magnitude limited sample reaches m$\rm_{5007}\leq 26.16$ mag, $\sim6$ magnitudes {fainter than} the bright cut-off for M 31. Figure~\ref{fig:pnlf_cum_disc} shows the completeness-corrected cumulative PNLFs for the PNe identified in the two disc regions of M~31. Figure~\ref{fig:pnlf_cum_subs} shows those computed for the six RGB overdensity regions. The different surface brightness and SFH of each substructure is reflected in the different total number of PNe observed in each region and in turn in the normalization  parameter, $c_1$, of the cumulative PNLF, with brighter regions showing a larger $c_1$ value than the fainter ones. The PNLF in each M 31 region is fitted by the generalised cumulative PNLF (in blue), and with $I_{\rm tot}(M)$ which includes the faint-end exponential (in orange). The fitted parameter values for all regions are tabulated in Table~\ref{table : pnlf}. For the N-Clump, the exponential function was unconstrained and hence its $c_{f1}$ and $c_{f2}$ values are not listed. 

We note that while $I_{\rm tot}(M)$ describes the cumulative PNLF of all the M~31 regions well, the fit to the cumulative PNLF of the 10<R$\rm_{GC} \leq $20 kpc disc region is less optimal in the m$\rm_{5007}\sim$ 20.5--21.5 mag interval. While the reason for this is presently unclear, our conjecture is as follows. This region has the largest number of PNe, N'$\rm_{PN}=624$, amongst all the regions studied here. Given the large number statistics, we may be sensitive to { sizeable dust attenuation in the 10<R$\rm_{GC} \leq $20 kpc disc region, as discussed in Section~\ref{sect:ext_sur}, and possibly to differences in the SFH likely to leave an imprint on the PNLF.}
{ Because of the variable dust attenuation in the inner disc, a fraction of the bright PNe would then be more extincted and thus occupy fainter magnitudes in the PNLF, thereby creating a slight paucity of PNe detected in the m$\rm_{5007}\sim$ 20.5--21.5 mag interval.}
Also it is possible that minor features in the cumulative PNLF, in the brighter magnitudes (m$\rm_{5007}\sim$ 20.5--21.5 mag), are not captured well by $I_{\rm tot}(M)$ and additional functions relating to other specific aspects of the SFH may be required. We discuss the measured effects of the parent stellar population on the M~31 cumulative PNLF in Sections~\ref{sect:pnlf_reg}~\&~\ref{sect:diff}. 

\begin{figure}[t]
	\centering
	\includegraphics[width=\columnwidth,angle=0]{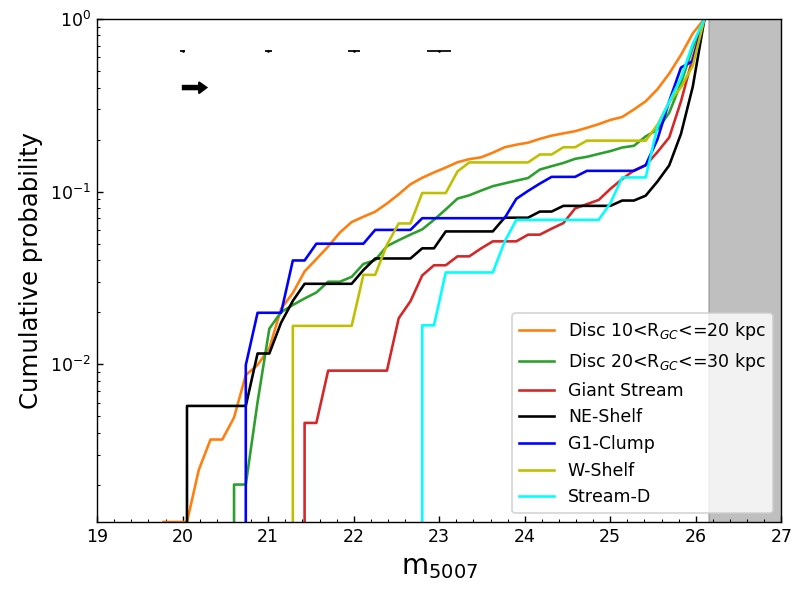}
	\caption{Cumulative probability of the completeness-corrected PNLF is shown (in log scale for visual clarity) for the five different substructures and the two disc regions. This diagnostic plot illustrates in which magnitude ranges the PNLFs are most different. The region fainter than the limiting magnitude of the shallowest field (m$\rm_{5007}$>26.16 mag) is shown in grey. In this Figure, we report the errorbars depicting the photometric errors (multiplied by 10 for visual clarity) with magnitudes in our survey. They are shown as black lines at m$\rm_{5007}=$ 20,21,22 and 23 mag respectively. The M* fitted values for each of the PNLFs are robust {in the entire survey area} given the photometric accuracy (m$\rm_{5007, err}\sim 0.014$ mag at m$\rm_{5007}=23$ mag). Furthermore, the arrow in the upper left corner shows the mean photometric extinction within our survey area (A$\rm_{5007}$=0.21 mag).}
	\label{fig:pnlf_cum_prob}
\end{figure}


\begin{table}[t]
\caption{Comparison of PNLF shapes for regions in M 31 with the Anderson-Darling test.}
\centering
\adjustbox{max width=\columnwidth}{
\begin{tabular}{c|c|c}
\hline
\multicolumn{2}{c|}{Regions compared} & Significance level (\%)\\
\hline
Disc (10<R$\rm_{GC} \leq $20 kpc) & \textit{Disc (20<R$\rm_{GC} \leq $30 kpc)} & \textit{4.1}\\
 & \textit{Giant Stream} & \textit{0.2}\\
 & \textit{NE-Shelf }& \textit{0.1}\\
 & \textit{G1-Clump} & \textit{0.1}\\
 & \textit{W-Shelf} & \textit{0.2}\\
 & \textit{Stream-D} & \textit{0.1}\\
\hdashline
Disc (20<R$\rm_{GC} \leq $30 kpc) & \textit{Giant Stream} & \textit{2.8}\\
 & \textit{NE-Shelf} & \textit{0.1}\\
 &\textit{ G1-Clump} & \textit{1.0}\\
 & \textit{W-Shelf} & \textit{0.2}\\
 & \textit{Stream-D} & \textit{0.1}\\
\hdashline
Giant Stream & NE-Shelf & 25.0\\
 & G1-Clump & 25.0\\
 & W-Shelf & 9.0\\
 & \textit{Stream-D} & \textit{0.1}\\
\hdashline
NE-Shelf & G1-Clump & 25.0\\
 & W-Shelf & 14.5\\
 & \textit{Stream-D} & \textit{0.7}\\
\hdashline
G1-Clump & W-Shelf & 25.0\\
 & \textit{Stream-D} & \textit{4.4}\\
\hdashline
W-Shelf & Stream-D & 18.1\\
\hline
\end{tabular}
\label{table : comp}
}
\tablefoot{The null hypothesis that the two PNLF shapes come from the same distribution is rejected at a given significance level. The value is floored / capped at 0.1\% / 25\%. In this work, we reject the null hypothesis when the significance level {is} less than 5\%. In such cases, the region {listed} in the second column and the {corresponding} significance level are marked in italics in the table.  {Origin of the differences in the PNLFs further discussed in Section~\ref{sect:disc} and Appendix~\ref{app:pair}.}}
\end{table}

\subsection{Comparison of the shapes of the PNLFs}
\label{sect:pnlf_comp}

The differences in the PNLFs obtained for distinct regions reflect diversities in their parent stellar populations (detailed in Section~\ref{sect:pnlf_reg}). We can thus establish the similarity or difference in stellar populations among substructures by statistically comparing their PNLFs, even if the color magnitude diagram for the resolved RGBs is not available for the entire region. We utilize the two-sample Anderson-Darling test \citep[AD-test;][]{ksample_ADtest} to compare PNLFs of distinct PN samples. For sample sizes larger than 8 \citep{lewis1961}, the two-sample AD-test is a non-parametric test which checks for the null hypothesis that two samples are extracted from the same distribution. The test statistic is computed from the distance between the cumulative probability distribution of the two samples. The test statistic thus obtained is compared against the critical values\footnote{In hypothesis testing, a critical value is a point on the test distribution that is compared to the test statistic to provide the significance level to reject the null hypothesis. For the AD-test, critical values have been defined by \citet{ksample_ADtest}.} to derive a significance level of the two samples being drawn from the same distribution. If the significance level is less than 5\%, the null hypothesis can be rejected and then the two samples are drawn from different distributions. Otherwise the two samples may or may not be drawn from the same distribution. The AD-test has high statistical power compared to other non-parametric tests like the Kolmogorov-Smirnov (KS) test, even for small sample sizes \citep{Razali11}. The AD-test is particularly more sensitive than the KS-test at differentiating between two distributions when the differences are most prominent near the beginning or end of the distributions. By comparing the PNLFs of distinct regions, we find which of these ones have statistically different parent stellar populations. 

The cumulative probability of the completeness-corrected PNLF, {computed by normalizing the PNLF between 0 and 1}, is shown for the distinct substructure regions in Figure~\ref{fig:pnlf_cum_prob}. We refrain from comparing the PNLF of the N-Clump with any other region because it has a sample size which is smaller than the minimum required for the AD-test. The significance level obtained for each compared pair of regions is tabulated in Table~\ref{table : comp}. 

We find that the PNLF of the disc regions are statistically different from each of the substructure regions. In many cases the significance level of two PNLFs being drawn from the same distribution is as low as 0.1\%.
We find that the two disc regions also have different PNLFs from each other {as their significance level of being drawn from the same distribution is 4.1\%}. We also find that the Stream-D has a PNLF, and hence a parent stellar population, different from the Giant-Stream, NE-Shelf and G1-Clump. From the AD-test results, the other substructure pairs (not marked in italics in Table~\ref{table : comp}) may or may not have different PNLFs, i.e. the test does not support a conclusive answer. If we recompute the AD-test on PN subsamples that are brighter than m$\rm_{5007}\le$25.9 mag, we find that the significance level of the AD-test for the two disc regions is 25\% (our capped value), meaning that their stellar populations may or may not be different. This indicates that the PNe in the brighter regions of the PNLF may originate from similar parent stellar populations, while any definite difference in stellar populations is engraved in the faint-end. Recomputing the AD-tests also for the substructures with PN subsamples brighter than m$\rm_{5007}=$25.9 mag, we find that all previous results are robust and not solely driven by the PNe in the faintest magnitude bin in the cumulative PNLFs, with the significance level well below 5\%. The exception is the comparison of the Disc (20<R$\rm_{GC} \leq $30 kpc) and Giant Stream regions, where the significance level increases to 9\%.

\subsection{{Quantitative differences of the PNLFs of distinct subregions in M31}}
\label{sect:obs_pnlf}

From the deep cumulative PNLFs extracted from the regions associated with the RGB overdensities and disc annuli in M31, we can assess in which magnitude ranges the cumulative PNLFs differentiate the most (see Figure~\ref{fig:pnlf_cum_prob}). 
\begin{itemize}
    \item  Bright-end of the PNLF: The apparent magnitude of the PNLF bright cut-off in the M~31 substructures falls in the range m$\rm_{5007}=$20--22.8 mag (Figure~\ref{fig:pnlf_cum_prob}) even though all these substructures are at the same LOS distance and have similar extinction values (see Section~\ref{sect:ext_sur} for details). In a given M31 subregion, any {variation of the fitted PNLF M$^*$ value (Table~\ref{table : pnlf})} from the reference $\rm M^{*}_{ref}= -4.54 \pm 0.05$ value, results from the intrinsic properties of that PN subsample and does not originate from a difference in LOS distance or extinction.
    \item Intermediate magnitudes of the PNLF: The $c_{2}$ parameter is the slope of the PNLF at intermediate magnitude (m$\rm_{5007}$= 21--25 mag). For some of the M~31 regions, we find $c_{2}\approx0$ (see Table~\ref{table : pnlf}). This occurs when there is a paucity of PNe at intermediate magnitudes and corresponds to a dip in the PNLF.  
    \item Faint-end of the PNLF: The rise in the faint-end of the PNLF is ubiquitous amongst the M31 region PNLFs as well as the LMC. However, the slope of the PNLF faint-end given by the $c_{f2}$ parameter covers a wide range of values (see Table~\ref{table : pnlf}) indicating differences at the faint-end of the PNLF.
\end{itemize}
Given the negligible effects of LOS distance, extinction, and the uniform/accurate photometry of our survey (Figure~\ref{fig:pnlf_cum_prob}), the differences in the PNLFs (as found in Section~\ref{sect:pnlf_comp}) are driven by the intrinsic properties of the PN subsamples and are linked to their parent stellar {populations}.


\section{Cumulative PNLFs and stellar populations in the M31 disc and inner halo}
\label{sect:pnlf_reg}
We can find possible correlations between the parameters that {describe} the difference among PNLFs of the M~31 regions with the metallicities ([M/H]; both photometric and spectroscopic as per availability) and SFH obtained from isochrone-fitting of colour-magnitude-diagrams (CMDs) of resolved stellar populations, mainly RGB stars, in these sub-regions and also for the LMC. 

\subsection{Variations of $M^*$ in the M31 subregions. Possible correlations with the metallicity of the parent population}
\label{sect:bright}

\begin{figure}[t]
	\centering
	\includegraphics[width=\columnwidth,angle=0]{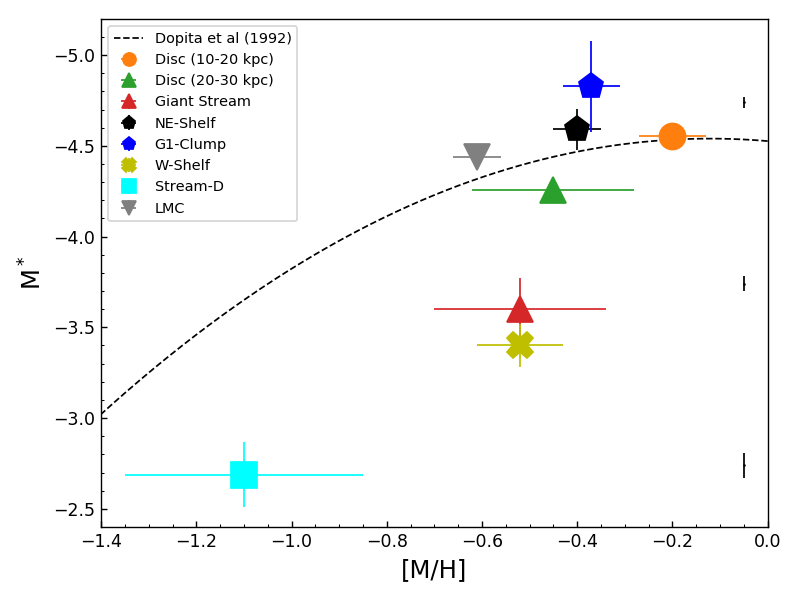}
	\caption{[M/H] for the different regions in M31 is plotted against the {$M^{*}$ value} obtained from fitting their PNLF. The different markers show the different references for the [M/H] measurements -- circle: \citet{wil17}; pentagon: \citet{bernard15}; triangle: \citet{Escala20}; inverted triangle: \citet{Nidever20}; cross: \citet{Tanaka10};  square:  \citet{Conn16}. The theoretical prediction for the values of {$M^{*}$} from \citet{Dopita92} models is shown with the dotted line. Also shown is the variation of photometric errors in our survey (multiplied by a factor $10$ for visual clarity) at the magnitudes corresponding to the given {$M^{*}$} values (as black lines at {$M^{*}$= -4.74, -3.74  and -2.74 mag}).}
	\label{fig:mstar}
\end{figure}

The inferred absolute magnitude of the PNLF bright cut-off M$^*$ in different galaxies was shown to be independent {of its morphology, star-formation history and age }\citep{ciardullo13}, with a weak dependency on the metallicity \citep[see][]{ciardullo02} for galaxies with $[M/H]$ larger than $-0.5$. In M31 we have the ideal laboratory to investigate this issue further as the PN subsamples and their PNLF are all at the same distance and with the same LOS extinction for radial distance $>20$ kpc from the center of M31.

{In Figure~\ref{fig:mstar}, we plot the $M^*$ for different M~31 regions against the tabulated [M/H] values of their parent stellar populations listed in Table~\ref{table : pnlf}. We provide the [M/H] values in different sub-regions in Appendix~\ref{app:met}, measured either from isochrone-fitting of the RGB-branch in CMDs or from spectroscopy of individual RGB stars which are available for some of the regions.} In Figure~\ref{fig:mstar}, we also depict the theoretical variation of {$M^*$} with [M/H]: such variation is computed from the dependency of $M^{*}$ on [O/H] obtained by \citet{ciardullo02} from the \citet{Dopita92} PN evolution models. [O/H] is calibrated to [M/H] by subtracting the solar [O/H] value \citep[=8.69;][]{Asplund09} as it is done for calibrating stellar and gas phase mass-metallicity relations of galaxies \citep[e.g.][]{Zahid17}. The theoretical dependence of $M^*$ values on $[M/H]$ predicts a dimming of $M^*$ with decreasing metallicity.

We find that the {$M^*$} for the LMC, the two disc regions, the G1-Clump and the NE-Shelf are in agreement with the theoretical prediction from \citet{Dopita92} PN models. However, the {$M^*$ values for the PN samples in the Giant Stream, W-Shelf and Stream-D areas are fainter than those predicted on the basis of the measured [M/H] of their parent stellar populations from the \citet{Dopita92} models.} We discuss hereafter in turn the possible origins of the deviant { $M^*$ }values.

Stream-D and W-Shelf substructures $-$ Only 5 and 10 PNe are brighter than m$\rm_{5007}=25$ mag in Stream-D and W-Shelf respectively, see Figure~\ref{fig:pnlf_cum_subs}. The low number of PN in these regions is related to the low luminosity of these substructures. {According to \citet{buz06} the visibility lifetimes of PNe are short plus the fading of a PN after the maximum brightness plateau is rather fast \citep{Schonberner07}. Hence in these M~31 substructures we may be preferentially detecting slow evolving fainter PNe from less massive, older stars that are entering the PN phase at their own dimmer $M^*$}. 

Giant Stream $-$ Differently from the previous two substructures, the Giant Stream is brighter and has a large number (113) of detected PNe.  Hence statistical uncertainty coupled with short PN visibility lifetime is not a viable alternative explanation for the fainter $M^*$ in the Giant Stream. The fainter value of $M^*$ for this substructure is further supported by the spectroscopic follow-up of its brightest PN. This PN (PN14 from \citealt{fang18}) has an absolute magnitude $M_{5007}=-3.51$ with $\mbox{12+(O/H)}= 8.3$ dex, corresponding to $[M/H]=-0.39$. Hence in Figure~\ref{fig:mstar} the {$M^*$} vs. [M/H] point derived from the PNLF fit and RGB metallicities \citep{Escala20} coincides with that from the $M_{5007}$ and [M/H] spectroscopic derived values for the brightest PN in the Giant Stream.

The current investigation indicates a significant discrepancy between the observed dimming of $M^*$ at $[M/H] \le -0.4$  and the more modest variations predicted from theoretical models of PN evolution at this [M/H] value. We note that the latter models from either \citet{Dopita92} or  \citet{Schonberner10} investigate the effect of varying metallicities only for the ionized nebula, while the central star is kept constant with solar metallicity. PN evolution models incorporating the effects of metallicity on the central star as well as on the ionized nebula may provide a better diagnostic for the effect of [M/H] on $M^{*}$. This is particularly important given the significance of the absolute magnitude of the PNLF bright cut-off, $M^{*}$, as a reliable secondary distance indicator across the different Hubble galaxy types \citep{ciardullo13}, including dwarf galaxies.

\subsection{Imprint of star-formation history on the faint-end of the PNLFs in M~31}
\label{sect:SFHonfaintpnlf}

\begin{figure}[t]
	\centering
	\includegraphics[width=\columnwidth,angle=0]{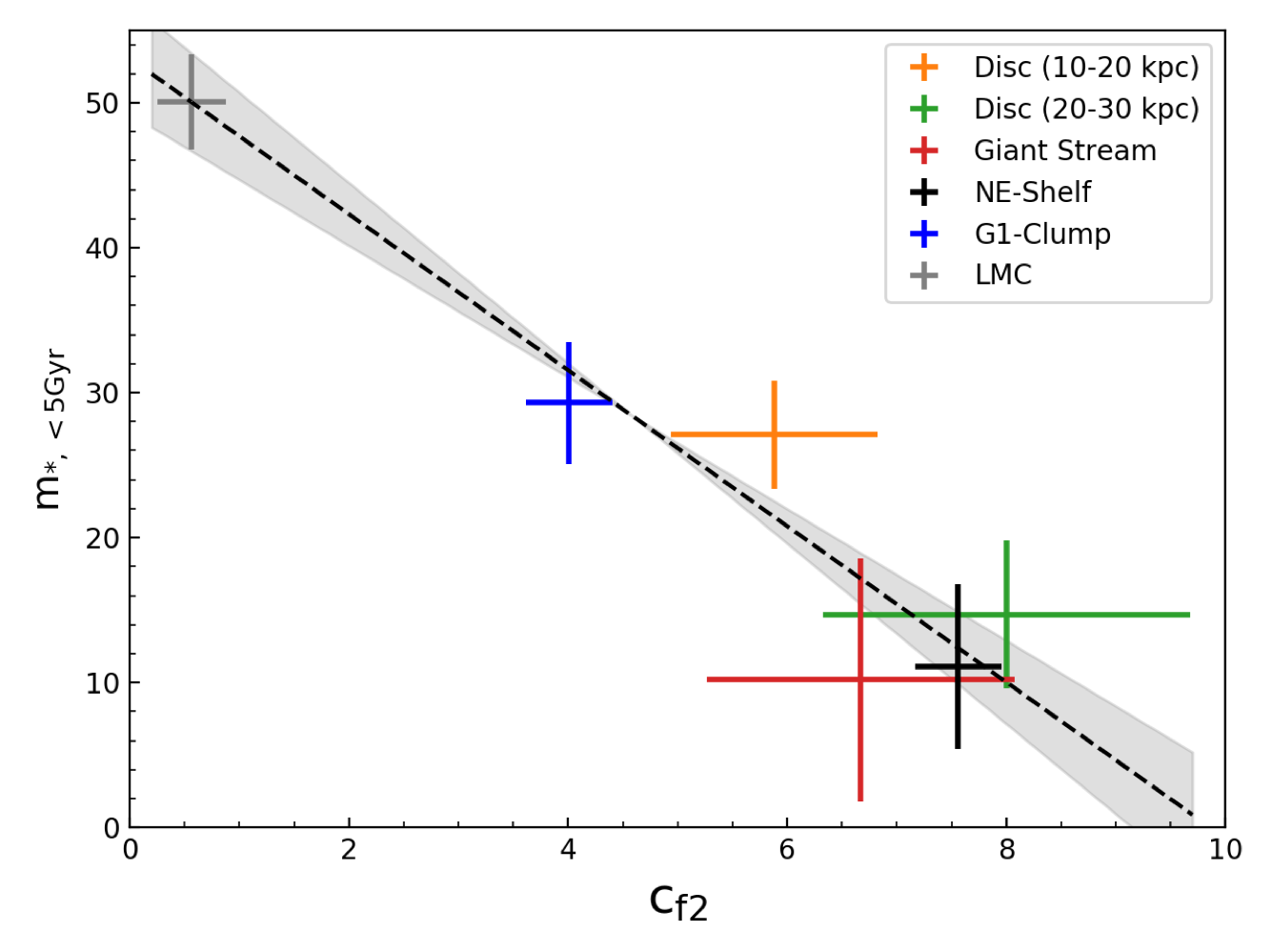}
	\caption{Linear relation between the slope of exponential function at the faint end of the PNLF, $c_{f2}$, and the percentage stellar mass formed within the last 5 Gyr,  $m\rm_{*,~<5~Gyr}$.}
	\label{fig:cf2_mass}
\end{figure}

\citet{bernard15} obtained the localized-SFHs in 14 deep HST pointings in different regions of the M~31 disc and some inner halo substructures. While they found that M~31 had a ubiquitous burst of star formation $\sim2$ Gyr ago, they measured significant differences in the percentage of stellar mass formed $<5$ Gyr ago (hereafter $m\rm_{*,~<5~Gyr}$) in distinct HST fields in different substructures. In particular, they found that at the location of the HST pointings in the Giant Stream, NE-Shelf and outer disc, the stellar populations have much smaller $m\rm_{*,~<5~Gyr}$ compared to the G1-Clump. Based on these two-age-bin SFHs ($<5$ and $>5$ Gyr), they termed the Giant Stream and NE-Shelf as {\it stream-like}, while the G1-Clump was termed as {\it disc-like}. The outer-disc field was labeled as a composite field, having a slightly larger  $m\rm_{*,~<5~Gyr}$ value compared to the {\it stream-like} fields. No SFH measurements are available for the W-Shelf\footnote{The HST field linked to the W-Shelf by \citet{bernard15} does not overlap with the W-Shelf spatial over-density as obtained by PAndAS \citep{mcc18}.} and Stream-D. \citet{wil17} utilized observations from the PHAT survey \citep{dal12} to find that $\sim10-20\%$ of the stellar mass within $R_{gc} = 20$ kpc is formed in a burst of star-formation $\sim2$ Gyr ago with  m$_{*, <5 Gyr}\sim30\%$. Using their HST observations in 8 scattered fields in the LMC, \citet{Weisz13} showed that $\sim50$ \% of the stellar mass in the LMC was formed in the last 5 Gyr. 

Given the extended coverage of the HST fields from which the SFH (from resolved stellar populations) in the LMC and M~31 10<R$\rm_{GC} \leq $20 kpc disc region are derived, the measured  $m\rm_{*,~<5~Gyr}$ values are representative of the average stellar population in these regions. Since the SFH of the average stellar population leaves an imprint on the global PNLF of any region, the measured   $m\rm_{*,~<5~Gyr}$ value for the aforementioned regions can be linked to their PNLF fit parameters. For the other four regions (see Figure~\ref{fig:spat}), the localized-SFH measured from the pencil-beam HST fields obtained by \citet{bernard15} may not be representative of the average stellar population for the entire substructures \citep{wil17}. 

We can assess whether the localized-SFH values determined from the pencil-beam HST fields are representative of the average values for the entire stellar population in any substructure by checking whether the relation between the global PNLF parameters and the measured $m\rm_{*,~<5~Gyr}$ value for these regions lie on the same relation as that for the SFH of the LMC and the M~31 10<R$\rm_{GC} \leq $20 kpc disc region, whose extended coverage results in measured SFHs from resolved stellar populations which are representative for the entire regions. Any region whose average stellar population is vastly different from that found within the tiny HST pointing would diverge significantly from any fitted relation. 

\begin{table}[t]
\centering
\caption{{Percentage stellar mass formed within the last 5 Gyr,  $m\rm_{*,~<5~Gyr}$, from the PNLF predicted with the pick-one-out test.}}
\adjustbox{max width=\columnwidth}{
\begin{tabular}{c|c|c}
\hline
Region & \multicolumn{2}{c}{$m\rm_{*,~<5~Gyr}$} \\
 & HST CMD & PNLF prediction\\
\hline
LMC & 50.1 $\pm$ 3.3 & 49.94 $\pm$ 11.98  \\
Disc (10-20 kpc) & 27.1 $\pm$ 3.7 & 20.02 $\pm$ 6.55  \\
Disc (20-30 kpc) & 14.7 $\pm$ 5.1 & 9.15 $\pm$ 11.02 \\
Giant Stream &  10.2 $\pm$ 8.4 & 18.04 $\pm$ 9.22 \\
NE-Shelf & 11.1 $\pm$ 5.7 & 13.49 $\pm$ 5.78 \\
G1-Clump & 29.3 $\pm$ 4.2 & 32.23 $\pm$ 2.77  \\
W-Shelf & N.A. & $\leq$ 42.19  \\
Stream-D &  N.A. & 27.14 $\pm$ 9.46  \\
\hline
\end{tabular}
\label{table : pred}
}
    \tablefoot{{The  $m\rm_{*,~<5~Gyr}$ in the HST fields corresponding to the LMC, Disc (10-20 kpc) and all other regions are from \citet{Weisz13}, \citet{wil17} and \citet{bernard15} respectively. The  $m\rm_{*,~<5~Gyr}$ values are predicted for each region from its $c_{f2}$ value by fitting the relation between $c_{f2}$ and $m\rm_{*,~<5~Gyr}$ for all the other regions except the selected region.}}
\end{table}

We find a correlation between the percentage stellar mass from the HST fields in two age bins (see Table~\ref{table : pred}), younger and older than 5 Gyr following the age distinction by \citet{bernard15}, and the exponential function fitting to the faint-end of the PNLF, $c_{f2}$ (see Table~\ref{table : pnlf}). Figure~\ref{fig:cf2_mass} shows $c_{f2}$ against $m\rm_{*,~<5~Gyr}$ in {the respective} HST field of a given sub-region. We find that the sub-regions whose SFH value is obtained from large spatial HST coverage lie tightly on the same relation as those regions with smaller fields. Fitting the relation between { $m\rm_{*,~<5~Gyr}$} and $c_{f2}$ with a linear function using a Deming regression\footnote{Regression technique where both variables are measured with error.} \citep{Kummell79}, the best fit is given by:
\begin{equation}
\label{eq:cf2}
    \rm m_{*,~<5~Gyr} = -5.38~(\pm 0.54)~\times~c_{f2}~+~ 53.07~(\pm 2.48)
\end{equation}

Since it is not known a priori the HST field of which region would be an outlier influencing the fitted parameters of Equation~\ref{eq:cf2}, we run a ``pick-one-out test'' over the parameters for the six regions. The relation between $c_{f2}$ and  $m\rm_{*,~<5~Gyr}$ is fitted for five of the six regions and the  $m\rm_{*,~<5~Gyr}$ value is predicted for the selected region from its $c_{f2}$ value. This process is iterated for each of the six regions. The predicted  $m\rm_{*,~<5~Gyr}$ values are noted for each region in Table~\ref{table : pred}. The predicted quantities are in good agreement with those measured from the HST fields implying that the $c_{f2}$~--~ $m\rm_{*,~<5~Gyr}$ relation is applicable to the entire parent stellar population of the PNe in any region. We thus also use Equation~\ref{eq:cf2} to predict the  $m\rm_{*,~<5~Gyr}$ values for the W-Shelf and Stream-D (Table~\ref{table : pred}).

\subsection{Stellar population dominating the very faint-end of the PNLF}
\label{sect:faint}

Following this analysis, we infer that the $c_{f2}$ value of the PNLF increases linearly with decreasing fraction of the stellar mass of its parent stellar population that formed in the last 5 Gyr. Hence, the rise at the faint-end of the PNLF is driven by the fraction of the stellar mass in the stellar population older than 5 Gyr. This is in agreement with the expectations from the stellar evolution models by \citet{Marigo04} where the faint-end of the PNLF is populated by PNe evolving from older stellar populations and powered by less-massive central stars (see their Figures 18 \& 25 and associated text). However, it is in contrast to the predictions by \citet{mendez08} where the faint-end of the PNLF is populated by PNe from massive progenitor stars from young stellar populations which have faded rapidly while losing their envelope.


\section{Discussion}
\label{sect:disc}

\subsection{Morphology of the PNLFs and stellar population parameters in the M~31 regions}
\label{sect:diff}

Through the analysis presented in Sections \ref{sect:pnlf} 
and \ref{sect:pnlf_reg}, we described our efforts to link the differences among the PNLFs of extended M~31 sub-regions to the diversity of the parameters describing their parent stellar populations. { We measured  different $M^*$ values of the PNLFs in the M~31 sub-regions. In case of the Giant Stream, W-Shelf and Stream-D, we can not link the measured dimmer $\rm M^*$ values of the PNLF to the lower metallicity of the parent stellar populations.} We further showed that the $c_{f2}$ values are clearly associated with the $m\rm_{*,~<5~Gyr}$. The $c_{2}$ parameter is also influenced by the SFH of the parent stellar population. \citet{Valenzuela19} showed theoretically that a dip in the PNLF is also a product of the distribution of masses of the central stars, and hence of the SFH of the parent stellar population. However, a quantitative link is yet to be established between the SFH and a dip in the PNLF, and hence its $c_{2}$ value.  

{For each pair of M 31 regions that have different PNLFs (see Table~\ref{table : comp}), we assessed in which magnitude ranges the PNLFs differentiate the most (Figure~\ref{fig:pnlf_cum_prob}) and summarized the differences in their fitted PNLF properties (see Table~\ref{table : pnlf}). We link the differences of pairs of M 31 regions to their parent stellar population parameters in Appendix~\ref{app:pair}. The highlights from this analysis are that the Giant Stream and the W-shelf have similarly faint $M^*$ values and large $c_{f2}$ values, with their mass in stars dominated by $>5$ Gyr old populations. Furthermore,} the NE-shelf has an $M^*$ value similar to the inner disc region within error, but its stellar mass is dominated by a $>5$ Gyr old population. The G1 clump has a $M^*$ value comparable to that of the inner disc region, with a significant contribution from a young $<5$ Gyr old population. {Finally}, Stream-D has the faintest $M^*$ value amongst the studied sub-regions.

\subsection{The merger origin of the inner-halo substructures in M~31}
\label{sect:merger}

Having mapped the average properties of the global stellar populations in the substructures of the M~31 inner halo, we now compare our observational results with the predictions from N-body/ hydrodynamical simulations of a merger event in M31. On one hand we have the N-body simulations of a minor merger (\citealt{Fardal13} and reference therein). This event, if it took place 1 Gyr ago, successfully reproduces the morphology of the Giant Stream, NE and W-shelves. The Giant Stream represents the trailing stream of material torn off during the progenitor’s first pericentric passage, while the NE and W shelf regions contain material torn off in the second and third passages, respectively \citep{fm16}. The largest fraction of stellar mass in all three substructures is made up by the old stellar debris of the satellite. This is consistent with the global PNLF results for these three sub-regions of M~31: the Giant Stream, and NE shelf have large fractions of older stars, i.e. large $c_{f2}$ values. The W-Shelf also has stellar populations different from the M~31 disc regions but the age of its stellar population is still not very well constrained.  

Such an evolution scenario, however, leaves out the formation of the N and G1 clumps. Following their initial discoveries \citep{Ferguson02,Ibata05}, these substructures had been linked to dissolution of dwarf galaxies from distinct accretion events. The G1-Clump is measured to have an absolute V-band Magnitude, M$_{V}=-12.6$ mag \citep{Ferguson02} which corresponds to a total V-band luminosity, L$_{tot}=9.37\times10^6$ M$_{\odot}$. Had the G1-Clump be the result of a dissolved dwarf, following the luminosity-metallicity relation for Local Group dwarf galaxies \citep{Kirby11}, its mean [M/H] value would be $\sim-1.5$. Such a value is much lower than its measured [M/H]$=-0.37$ (Table~\ref{table : pnlf}). Thus, the G1-Clump, given its relatively large metallicity, could not have formed from the dissolution of a small dwarf.

The remaining issues regarding the minor merger scenario are the tension between the timing of such an event, $\sim 1$ Gyr ago, and the age of the burst of star formation of $\sim 2$ Gyr in the M31 disc \citep{bernard15,wil17}, and particularly the mass of the satellite not being large enough to dynamically heat the M31 disc \citepalias{Bh+19b}. 

We now examine the predictions from the major-merger scenario as simulated by \citet{ham18}. In these simulations, a massive gas-rich satellite was accreted with an orbit along the Giant Stream 2-3 Gyr ago. Such a massive satellite would perturb the M~31 disc and produce a thick disc from the pre-existing stars. After the merger, the replenished cold gas would lead to a burst of star-formation and the build-up of a less-extended thin disc. Within the 10<R$\rm_{GC} \leq $20 kpc disc region, we observe in the PNLF a superposition of the stellar populations associated with both the newly-formed thin disc and the older thicker disc of M~31. This disc region thus has a larger fraction of younger stars compared to the 20<R$\rm_{GC} \leq $30 kpc disc region where the stellar populations associated with the thicker disc stars dominate.

The simulations by \citet{ham18} also predict that the stellar populations in the Giant Stream, NE-Shelf and W-Shelf are dominated by stellar debris from the infalling satellite, while the N and G1 Clumps are associated with the stellar material from the perturbed pre-existing disc. The G1 Clump has a significant fraction of stars younger than 5 Gyr just like the thin disc of M~31. While their stellar populations are statistically different, the younger stars in both the thin disc of M~31 and the G1 Clump may have formed at the same time.  Having formed from the perturbed M~31 disc, the PNLF of the G1-Clump is expected to have a $M^*$ value similar to the inner disc and a slope in the faint end consistent with significant contribution from a young stellar populations,  in line with the observed values. Unfortunately, the stellar population of the N-Clump could not be constrained by our data. The major merger scenario thus simultaneously explains the observed global stellar population properties of the M~31 inner halo substructures, as well as the measured velocity dispersion values \citepalias{Bh+19b} and the age of the recent burst of star-formation in the disc \citep{bernard15,wil17}. 

Stream-D has a stellar population that is different from both disc regions and the other substructures: it has the faintest value for the PNLF bright cut-off $M^*$ and its stellar population is the most metal poor \citep{Conn16} in the entire area surveyed around M31. This result points to an independent origin. Indeed this stream does not appear in the simulations by \citet{ham18} and it could have formed in a distinct accretion event, e.g. the disruption of a low mass dwarf galaxy. Thus we find that the M~31 inner halo substructures, barring Stream-D, are consistent with having originated in a single major merger event.

\section{Conclusions}
\label{sect:future}

{This study investigates the properties of the PNLF for the stellar populations in different subregions of M31, based on the largest PN sample ever obtained in any galaxy, including the Milky Way. Because of the unprecedented depth, spatial coverage and uniform photometry of the survey we can empirically investigate the variations of the PNLF across different stellar populations with robust statistics, including the absolute magnitude $M^*$ of the bright cut-off, and the faint-end slope $c_{f2}$. This is greatly aided by the fact that all probed PN sub-populations are at similar distance, and beyond $R>20$ kpc from M31's centre have constant LOS extinction.} Furthermore, for the first time we can use the properties of the PNLF to characterize the average stellar populations of the different substructures in M31 and thereby constrain the merger formation history of the galaxy's inner halo. {Our main results are as follows:}
\begin{enumerate}
    \item {Statistical comparison of the deep PNLFs in different substructures and disc annuli of M~31 reveal significant differences in the fitted PNLF parameters. In particular $M^*$, the PNLF bright cut-off, is found to vary strongly: for the Giant Stream and W-shelf,  $M^*$ is nearly one magnitude fainter than for the disc, and the low-metallicity stream D has even fainter  $M^*$.}
    \item {The $M^*$ values of the PNLFs for the LMC, NE-Shelf, G1-Clump and the two M~31 disc regions are consistent with the theoretical predictions by \citet{Dopita92} of $\rm M^*$ dimming in more metal poor populations (see Section~\ref{sect:bright}). However, the measured $M^*$ value for the Giant Stream, W-Shelf and Stream-D are much larger than theoretical predicted. The empirically measured variations of $M^*$ in the sub-regions of M~31 and the LMC can thus not be conclusively attributed to variations only in [M/H], thereby highlighting the need for theoretical models of PN evolution with improved treatment for metallicity and other possible effects such as the parent stellar population's age.}
    \item {Using published HST data, we establish a connection between the PNLF morphology and the stellar mass of the parent stellar population in two age bins, younger and older than 5 Gyr, respectively. We find that the rise in the faint-end of the PNLF is driven by the mass-fraction of stellar populations older than 5 Gyr, indicating that PNe evolving from older stars populate the faint-end of the PNLF.}
    \item {From their PNLFs we find that the Giant Stream and NE-Shelf are consistent with being composed of stellar debris from an in-falling satellite, while the G1 Clump is linked to the pre-merger M~31 disc with a significant contribution of younger stars presumably formed out of the accreted cold gas.}
    \item {The constraints from the deep PNLF on the stellar populations of these substructures point to their origin in a single merger event in M~31. In conjunction with the age-velocity dispersion relation of the M~31 disc \citepalias{Bh+19b}, this indicates a recent ($\sim2$-$3$ Gyr ago), fairly massive merger event in M~31, in agreement with predictions from \citet{ham18}. We also find that Stream-D was likely formed in a distinct accretion event, probably from a disrupted dwarf satellite.}
\end{enumerate}

This work sets the stage for the spectroscopic follow-up of PNe in the outer disc and inner halo of M31, with the goal of measuring their specific metal abundances and accurate LOS velocities. From these discrete measurements, we will then build abundance maps for substructures and disc, correlate variations of metal abundance with kinematics signatures, thereby leading to a better understanding of their chemical and dynamical evolution.


\begin{acknowledgements}
      Based on observations obtained with MegaPrime/ MegaCam, a joint project of CFHT and CEA/DAPNIA, at the Canada-France-Hawaii Telescope (CFHT) which is operated by the National Research Council (NRC) of Canada, the Institut National des Sciences de l'Univers of the Centre National de la Recherche Scientifique of France, and the University of Hawaii. We acknowledge comments and suggestions from K. Gesicki.  { We thank the A\&A Editor, M. Salaris, for a constructive revision process.} SB acknowledges support from the IMPRS on Astrophysics at the LMU Munich. SB, MA and OG are grateful for the hospitality of the Mount Stromlo Observatory and the Australian National University (ANU). MA and OG thank the Research School of Astronomy and Astrophysics at ANU for support through their Distinguished Visitor Program. This work was supported by the DAAD under the Australia-Germany joint research program with funds from the German Federal Ministry for Education and Research. This research made use of Astropy-- a community-developed core Python package for Astronomy \citep{Rob13}, SciPy \citep{Scipy}, NumPy \citep{numpy} and Matplotlib \citep{matplotlib}. This research also made use of NASA’s Astrophysics Data System (ADS\footnote{\url{https://ui.adsabs.harvard.edu}}).
\end{acknowledgements}


\bibliographystyle{aa} 
\bibliography{ref_pne.bib}

\begin{appendix} 

\section{PN circumstellar extinction and the PNLF}
\label{sect:ext}

A fraction of the dust produced in stars during the Asymptotic Giant Branch (AGB) phase \citep[e.g.][]{Ventura14} is retained in the PN phase of stellar evolution. This dust production in the AGB phase depends on its initial stellar mass, and hence on the SFH of its parent stellar population \citep[e.g.][see their Fig 10]{Ventura14}. In fact, \citet{Ciardullo99} found that PNe with more massive progenitors have higher measured circumstellar extinction. \citet{davis18} found that the M$^*$ value of the PNLF in the M~31 bulge is set from a combination of the [\ion{O}{III}] 5007\AA\ flux emitted by the nebula and its circumstellar extinction. PNe with high [\ion{O}{III}] 5007\AA\ flux also have high circumstellar extinction such that the net observed [\ion{O}{III}]\ 5007\AA\ flux keeps the PN fainter than the $M^*$ value of the PNLF (see their Figure 13). In case the de-reddened [\ion{O}{III}] 5007\AA\ flux is considered, the resultant de-reddened PNLF (see their Figure 14) does not follow the generalised PNLF functions \citep{ciardullo89,longobardi13}. Thus the SFH dependent circumstellar extinction (a property of individual PN) should not be included in obtaining the PNLF and its $M^*$ value for any of the regions. 

Furthermore, the brightest PN in the Giant Stream \citep[PN14;][]{fang18} has a extinction (LOS plus circumstellar) of 0.37 mag (spectroscopically measured from its Balmer decrement), much lower than the observed difference between the Giant Stream $M^*$ value and the predicted value by \citet{Dopita92} for its [M/H] value (see Figure~\ref{fig:mstar}). This further confirms that the any difference in the M$^*$ values of the M31 region PNLFs does not result from different extinctions, even if circumstellar extinction is taken into account.

\section{[M/H] measurements for the LMC and M~31 regions}
\label{app:met}
In general, the photometrically measured [Fe/H] is taken to be equal to the [M/H] value, as [$\rm\alpha$/Fe] is assumed to be zero for the fitted isochrones \citep[e.g.][]{bernard15,Conn16,wil17}.  The [M/H] for the 10<R$\rm_{GC} \leq $20 kpc disc region is obtained as the mean [M/H] from the PHAT photometry in the radial range R$\rm_{GC}=$ 12--20 kpc, covering a third of the M~31 disc at this radii, by \citet{wil17}. The [M/H] for the the NE-Shelf and G1-Clump regions is from the pencil-beam HST photometry by \citet{bernard15}, with the errors obtained from their metallicity distribution functions. For Stream-D, the [M/H] measurement is from the CFHT MegaCam observations, covering this entire sub-region, by \citet{Conn16}. The [M/H] for the 20<R$\rm_{GC} \leq $30 kpc disc region and the Giant Stream \footnote{The metallicity gradient in the Giant stream is studied by \citet{Conn16}. In their Figure 3, they show the spatial coverage of their fields and give the corresponding [Fe/H] (=[M/H] as [$\alpha$/Fe]=0) values in their Table 1. With this PNe survey, we cover a spatial area in the Giant Stream  corresponding to their mean metallicity range of [M/H] = $-0.4$ to $-0.7$. The [M/H] value  for the Giant Stream from \citet{Escala20} covers the metallicity range observed by \citet{Conn16} within error and is thus an accurate [M/H] value for parent population of the PN subsample in the Giant Stream.} are obtained from the mean [Fe/H] and mean [$\rm\alpha$/Fe] values measured from individual stars in small fields observed in these regions by \citet{Escala20}. These values are converted to [M/H] using the following relation from \citet{Salaris05}:
\begin{equation}
\label{eq:mh}
    \rm [M/H] \sim [Fe/H] + log(0.694\times10^{[\alpha/Fe]}+0.306)
\end{equation}
We note that the spectroscopic [M/H] for these two regions agree within errors with that from the pencil-beam HST photometry in these regions by \citet{bernard15}. For the LMC, we derive the median [M/H] value from the median [Fe/H] and median [$\rm\alpha$/Fe] values for APOGEE RGB stars spanning a large radial range in the LMC \citep{Nidever20}. For the W-Shelf, \citet{Tanaka10} construct CMDs for the resolved stellar population in the Subaru Supreme-Cam pointings. The mean [Fe/H] value is measured photometrically from isochrone-fitting to the W-Shelf CMD, assuming [$\rm\alpha$/Fe]=0.3. We use the measured [Fe/H] and assumed [$\rm\alpha$/Fe] value in Equation~\ref{eq:mh} to obtain the measured [M/H] for the W-Shelf.

\section{Pair-wise differences in the PNLFs of M~31 regions and their stellar population parameters}
\label{app:pair}
{For the pairs of M~31 regions that have different PNLFs (see Table~\ref{table : comp}), we assessed in which magnitude ranges the PNLFs differentiate the most (Figure~\ref{fig:pnlf_cum_prob}) and presented the differences in their fitted PNLF properties (see Table~\ref{table : pnlf}). We infer the dependencies of these differences in pairs of M~31 regions on their parent stellar population parameters as follows: 
\begin{itemize}
\item The 10< R$\rm_{GC}~\leq $20 kpc and 20< R$\rm_{GC}~\leq $30 kpc disc regions: On the basis of the spectroscopic and kinematic properties of the disc PNe \citepalias{Bh+19b}, the 10<R$\rm_{GC}~\leq $20 kpc disc region, which contains both young (heated) thin disc and older thicker disc stars, is found to be different from the 20<R$\rm_{GC} \leq $30 kpc disc region, in that the latter has predominantly older and more metal poor disc stars, reflected in its larger $c_{f2}$ (see Table~\ref{table : pnlf}) and fainter $M^*$ values respectively.
\item The Giant Stream and the 10<R$\rm_{GC}~\leq $20 kpc disc region: The Giant Stream, with a larger $c_{f2}$ value (see Table~\ref{table : pnlf}), has a higher percentage of older stars than the 10<R$\rm_{GC}~\leq $20 kpc disc region. Additionally, the PNLF of the Giant Stream has a fainter M$^*$ (see Section~\ref{sect:bright}) than the 10<R$\rm_{GC} \leq $20 kpc disc.
\item The Giant Stream and 20<R$\rm_{GC} \leq $30 kpc disc region:  Here the difference in the deep PNLFs stems mainly from the fainter M$^*$ (see Table~\ref{table : pnlf}) of the Giant Stream PNLF with respect to that of the outer disc region. Their stellar mass is dominated by older stars according to their large $c_{f2}$ values.
\item The NE-Shelf and 10<R$\rm_{GC} \leq $20 kpc disc region: The NE-Shelf has a higher $c_{f2}$ value (see Table~\ref{table : pnlf}), thus a higher percentage of older stars, than the 10<R$\rm_{GC} \leq $20 kpc disc region. While the NE-shelf PNLF has a $M^*$ value similar to that of the inner disc region within error, it has a $c_{2}\approx0$ value (see Table~\ref{table : pnlf}), indicating further differences in the SFH from the 10<R$\rm_{GC} \leq $20 kpc disc region.
\item The NE-Shelf and 20<R$\rm_{GC} \leq $30 kpc disc region: The difference between the two regions originates from the slightly brighter $M^*$ of the NE-Shelf PNLF, and also the latter having a $c_{2}\approx0$ value, implying differences in their SFHs.
\item The G1-Clump and the 10<R$\rm_{GC} \leq $20 kpc disc region: Both regions have similar {\it $m\rm_{*,~<5~Gyr}$} (see Table~\ref{table : pred}) but the PNLFs have a different shape owing to the $c_{2}\approx0$ value (see Table~\ref{table : pnlf}) for the G1-Clump, thus indicating a distinct SFH from the 10<R$\rm_{GC} \leq $20 kpc disc region. 
\item The G1-Clump and the 20<R$\rm_{GC} \leq $30 kpc disc region: The G1-Clump has a higher {\it $m\rm_{*,~<5~Gyr}$} (see Table~\ref{table : pred}) and lower $c_{f2}$ value (see Table~\ref{table : pnlf}) than the 20<R$\rm_{GC} \leq $30 kpc disc. The difference is further aggravated due to the $c_{2}\approx0$ value of the G1-Clump PNLF, implying differences in SFHs.
\item The W-Shelf and the two disc regions: We only have an upper limit of {\it m$_{*, <5 Gyr}\leq$}42.19 \% for the W-Shelf. The difference with the two disc regions can still be attributed to the $c_{2}\approx0$ value of the W-Shelf PNLF (see Table~\ref{table : pnlf}) and thus a difference in SFHs. Additionally, the W-Shelf PNLF has a fainter M$^*$ (see Section~\ref{sect:bright}) than the two disc regions.
\item Stream-D: This M~31 substructure is different from both disc regions, the Giant Stream, the NE shelf and the G1-Clump  because it has the faintest bright cut-off value, $M^{*}=-2.689 \pm 0.177$, among the PN subsamples from the M31 survey (see Section~\ref{sect:bright}). Its deep PNLF also has $c_{2}\approx0$ (see Table~\ref{table : pnlf}), further implying distinct SFHs to some of the M~31 sub-regions.
\end{itemize}}
 {A summary of the highlights is provided in Section~\ref{sect:diff}}.
\end{appendix}
\end{document}